\documentclass[reprint, superscriptaddress, secnumarabic, amssymb, nobibnotes, aps, prl]{revtex4-1}

\usepackage{graphicx}
\usepackage{epstopdf}
\usepackage[T1]{fontenc}
\usepackage[latin9]{inputenc}
\usepackage{amsbsy}
\usepackage{gensymb}
\usepackage[T1]{fontenc}
\usepackage[latin9]{inputenc}
\usepackage{amsmath}
\usepackage{amssymb}
\usepackage{bbm}
\usepackage{braket}
\usepackage{xcolor}
\allowdisplaybreaks
\usepackage{graphicx}
\usepackage[colorlinks=true]{hyperref}  
\hypersetup{
    bookmarks=true,         % show bookmarks bar?
    unicode=false,          % non-Latin characters 
    pdftoolbar=true,        % show Acrobat
    pdfmenubar=true,        % show Acrobat 
    pdffitwindow=false,     % window fit to page when opened
    pdfstartview={FitH},    % fits the width of the page to the window
    pdftitle={x},    % title
    pdfauthor={SM},     % author
    pdfsubject={},   % subject of the document
    pdfcreator={},   % creator of the document
    pdfproducer={}, % producer of the document
    pdfkeywords={} {} {}, % list of keywords
    pdfnewwindow=true,      % links in new window
    colorlinks=true,       % false: boxed links; true: colored links
    linkcolor=blue, %red,          % color of internal links (change box color with linkbordercolor)
    citecolor=blue,        % color of links to bibliography
    filecolor=magenta,      % color of file links
    urlcolor=blue           % color of external links
} 
\usepackage[normalem]{ulem}

\renewcommand{\approx}{\simeq}

\begin{document}
\title{\textrm{Magnetism and exchange bias properties in Ba$_{2}$ScRuO$_{6}$}}
\author{Prachi Mohanty}
\affiliation{Indian Institute of Science Education and Research Bhopal, Bhopal, 462066, India}
\author{Sourav Marik}
\email[]{soumarik@thapar.edu}
\affiliation{School of Physics and Materials Science, Thapar Institute of Engineering and Technology, Patiala 147004, India}
\author{R. P. Singh}
\email[]{rpsingh@iiserb.ac.in}
\affiliation{Indian Institute of Science Education and Research Bhopal, Bhopal, 462066, India}

\date{\today}
\begin{abstract}
\begin{flushleft}

\end{flushleft}
This paper presents structural, detailed magnetic, and exchange bias studies in polycrystalline Ba$_{2}$ScRuO$_{6}$ synthesized at ambient pressure. In contrast to its strontium analogue, this material crystallizes in a 6L hexagonal structure with the space group P$\overline{3}$m1. The Rietveld refinement using the room-temperature powder X-ray diffraction pattern suggests a Ru-Sc disorder in the structure. The temperature variation of the dc-electrical resistivity highlights a semiconducting behaviour with the electron conduction corresponding to the Mott 3D-VRH model. Detailed magnetization measurements show that Ba$_{2}$ScRuO$_{6}$ develops antiferromagnetic ordering at T$_{N}$ $\approx$ 9 K. Interestingly, below 9 K (T$_{N}$), the field cooled (FC) magnetic field variation of the magnetization curves highlights exchange bias effect in the sample. The exchange bias field reaches a maximum value of 1.24 kOe at 2 K. The exchange bias effect below the magnetic ordering temperature can be attributed to inhomogeneous magnetic correlations owing to the disorder in the structure. 
\end{abstract}
\maketitle
\section{Introduction}

 Double perovskites (DP), discovered in 1889, with a general formula A$_{2}$BB$'$O$_{6}$ (A = Alkaline metals, B and B$'$ = transition metals) have attracted intense research attention, as these materials are known to host intriguing magnetic and magneto-transport properties that have potential applications in spintronics \cite{1a,2a,3a,4a,5a,6a}. The two transition metal sites (B and B$'$) form interpenetrating face-centered cubic (FCC) sublattices in an ideal DP structure. As a structural feature of DPs, the intra and inter-sublattice superexchange couplings and their interactions determine the complex magnetic properties of the system \cite{19a,20a,21a,22a}. However, the double perovskites with magnetic ion in the B$'$ site (B = nonmagnetic) attract recent interest due to their potential to design an enhanced magnetic frustration and exchange bias effect \cite{15s,19s}. 

Among double perovskites, ruthenium-based DPs with composition A$_{2}$BRuO$_{6}$ (B = smaller metals) is an important family of DPs due to their rich structural variation and diverse range of exotic magnetic behavior \cite{7g, 8g, 9g, 10g, 11g, 12g, 13g, 14g, 15g, 16g, 17g, 18g, 19g, 19gg, 19gg1, 19gg2}. The structure has two nonequivalent crystallographic sites for Ru and B-type cations. This arrangement is similar to a geometrically frustrated edge-shared tetrahedra network; therefore, the antiferromagnetic (AF) correlations are frustrated. However, several Ru$^{5+}$ DPs show exotic magnetic behavior and long-range antiferromagnetic ordering despite high magnetic frustration. For instance, Ba$_{2}$YRuO$_{6}$ (cubic) \cite{15b} shows a very high Curie-Weiss temperature (-522 K); however, a long-range magnetic ordering is observed at 37 K, Sr$_{2}$YRuO$_{6}$ shows magnetization reversal and exchange bias-like properties \cite{13b,17bb,17b}. Monoclinic La$_{2}$LiRuO$_{6}$ and La$_{2}$NaRuO$_{6}$ also show long-range magnetic ordering despite high magnetic frustration \cite{17s, 17ss}. In addition to the interesting half-filled configuration of Ru$^{5+}$ (t$_{2g}$$^{3}$, S = 3/2) cations, the role of spin-orbit coupling (SOC) is also crucial in determining the magnetic and physical properties of the 4d (and 5d) containing materials \cite{17ss1}. Furthermore, the presence of non-magnetic B cation in the structure can affect the strength of the magnetic correlations.  

Recently, Kayser et al. \cite{BaSc} highlighted a new Ru-Sc-based double perovskite. Their study shows that the ambient pressure synthesized Ba$_{2}$ScRuO$_{6}$ can be stabilized in a 6L hexagonal structure. Interestingly, a disorder of Ru - Sc is observed in the structure. The inhomogeneous distribution of magnetic and non-magnetic cations in different crystallographic sites can trigger exotic magnetic behavior, such as frustrated magnetism and the exchange bias effect.  
Here, we report the detailed magnetic properties and exchange bias effect in a ruthenium DP compound with composition Ba$_{2}$ScRuO$_{6}$.   
\section{Experimental Section}

A standard solid state reaction method is used to synthesize the polycrystalline Ba$_{2}$ScRuO$_{6}$ material. Stoichiometric amounts of BaCO$_{3}$, Sc$_{2}$O$_{3}$ and RuO$_{2}$ were homogenized using a mortar and pastel. The sample was heat treated at 960$^{\circ}$C and then at 1400$^{\circ}$C for 72 h in an air atmosphere. Each step involves several intermediate grindings and pelleting. Powder x-ray diffraction (XRD) patterns were collected at room temperature (RT) using a Panalytical X'pert Pro diffractometer (Cu K$\alpha$-radiation, $\lambda$ = 1.5406$\AA$). Phase purity and structural analysis were done by Rietveld analysis using the RT - XRD patterns (using Full Prof suite software). Electrical resistivity ($\rho$) data as a function of temperature were collected in the absence of a magnetic field using a Physical Property Measurement System (PPMS, Quantum Design). The temperature-dependent resistivity measurement was performed using a four-probe dc method. Magnetization measurements as a function of temperature and magnetic field were performed using a Quantum Design MPMS-3 magnetometer. We have collected magnetization data in the temperature range of 1.9 K $<$ T $<$ 400 K. The magnetic field variation of the magnetization (M-H) loops were measured between $\pm$ 2.5 kOe.

\begin{figure}[h!]
\begin{center}
\includegraphics[scale=0.5]{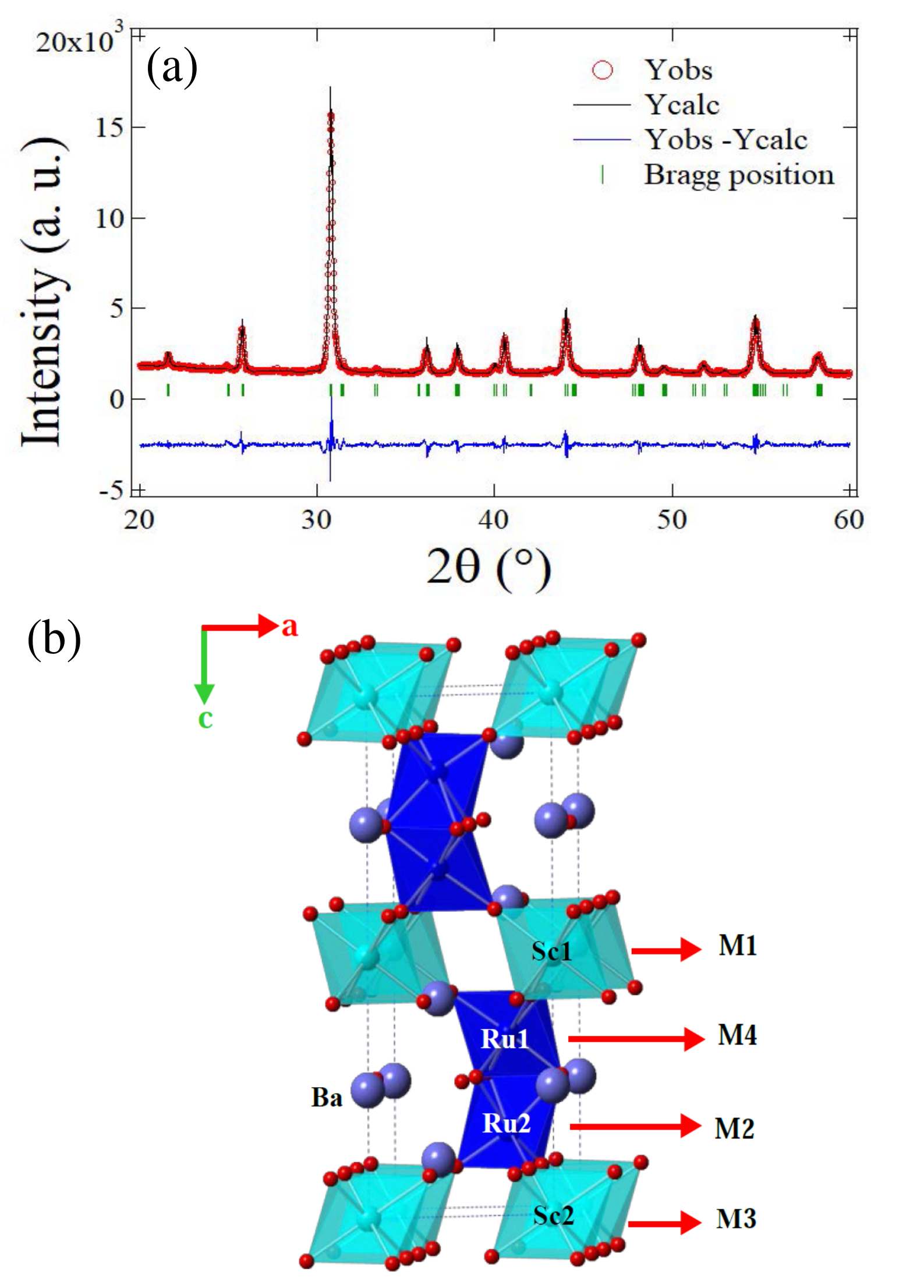}
\caption{\label{fig 1}  (a) Rietveld refinement plot of the X-ray powder diffraction pattern obtained at room temperature and (b) the hexagonal perovskite-type structure representation (space group P$\overline{3}$m1) for Ba$_2$ScRuO$_6$. The Ba atoms are shown as blue spheres. Oxygen atoms are highlighted as red spheres. The center of the polyhedra is occupied by Sc and Ru cations. Ru occupies the M2 and M4 sites (face-shared) primarily. Scandium occupies the corner-shared M1 and M3 sites.}
\end{center}
\end{figure}

\section{Results and discussion}
 \begin{table}[h!]
\caption{Structural parameters for Ba$_{2}$ScRuO$_{6}$ at RT extracted from Rietveld refinement of the powder XRD diffraction data}
\label{BaSc}
\begin{center}
~~~~~~~~~Structure~~~~~~Hexagonal\\
 ~~~Space group~~P$\overline{3}$m1\\
~~~~~~~~~~~~~\textit{a}(\AA) = 5.80946(3)\\
~~~~~~~~~~~~~\textit{b}(\AA) = 5.80946(3)\\
~~~~~~~~~~~~~\textit{c}(\AA) = 14.3028(3)\\
~~~~~~~\textit{$\alpha$} = 90\\
~~~~~~~\textit{$\beta$} = 90\\
~~~~~~~\textit{$\gamma$} = 120\\
~V$_{cell}$(\AA$^{3}$) = 418.04\\
\vspace{5mm}
\begin{tabular}[b]{lccc}\hline\hline
Atom & x & y & z\\
\hline
%\\[0.5ex]                                  
Ba1  & 0 & 0 & 0.25\\             
Ba2  & 0.33 & 0.67 & 0.102\\
Ba3  &  0.33 & 0.67 & 0.413\\                       
Sc1 (M1)  & 0 & 0 &  0.5\\
Sc2 (M3)  & 0 & 0 &  0\\
Ru1 (M2)  & 0.33 & 0.67 & 0.661\\
Ru2 (M4)  & 0.33 & 0.67 & 0.848\\
O1  & 0.162 & -0.162 & 0.299\\
O2  & 0.820 & -0.820 & 0.081\\
O3  & 0.835 & -0.835 & 0.415\\
%\\[0.5ex]
\hline\hline
\end{tabular}
\par\medskip\footnotesize
\end{center}
\end{table}
Phase purity and crystal structure were determined using room-temperature (RT) X-ray diffraction (XRD). The Rietveld refinement of the room temperature X-ray diffraction pattern (Fig.\ref{fig 1}) highlights that Ba$_{2}$ScRuO$_{6}$ crystallizes in a single
phase hexagonal symmetry, space group (S.G.) P$\overline{3}$m1 with the Sc and Ru ordering in alternate (001) layers.  The final results (cell parameters and atomic positions) of the structural analysis using the RT-XRD pattern are listed in Table 1. The obtained lattice parameters  are a = b = 5.80946(3) \AA,  and c = 14.3028 (3) \AA, and are in good agreement with the previous report \cite{BaSc}. Four different crystallographic sites (Sc1 (1b, 0, 0, 0.5), Ru1 (2d, 0.33, 0.67, z), Sc2 (1a, 0, 0, 0), Ru2 (2d, 0.33, 0.67, z)) are used to describe the positions of Sc and Ru cations (in the crystal structure as shown in Fig.\ref{fig 1} (b), they are depicted as M1 (1b), M4 (2d), M3 (1a) and M4 (2d)). As suggested by Kayser et al. \cite{BaSc}, a disorder of Ru - Sc is considered in our refinement. The Ru-Sc disorder is observed at the crystallographic sites of Sc2 (M3, 1a) and Ru2 (M4, 2d). However, the other two sites, M and M2, are filled by the Sc (M1, 1b) and Ru (M2, 2d) cations, respectively.  The occupancy refinements of Ba show full occupancy. The positions of the oxygen atoms in the structure are described using three different oxygen positions (table 1). However, the occupancy of the oxygen positions was fixed during the refinement. A view of the crystal structure for Ba$_{2}$ScRuO$_{6}$ material is depicted in the Fig.\ref{fig 1} (b).  The structure involves the face sharing M$_{2}$O$_{9}$ dimeric units linked by a single corner-sharing MO$_{6}$ octahedral units. The octahedra with Sc atoms at the center constitute sets of corner-shared MO$_{6}$ units. However, Ru cations occupy face-sharing positions and connect to the Sc-centered octahedra by corner-sharing.

\begin{figure}[h!]
\begin{center}
\includegraphics[width=1\columnwidth]{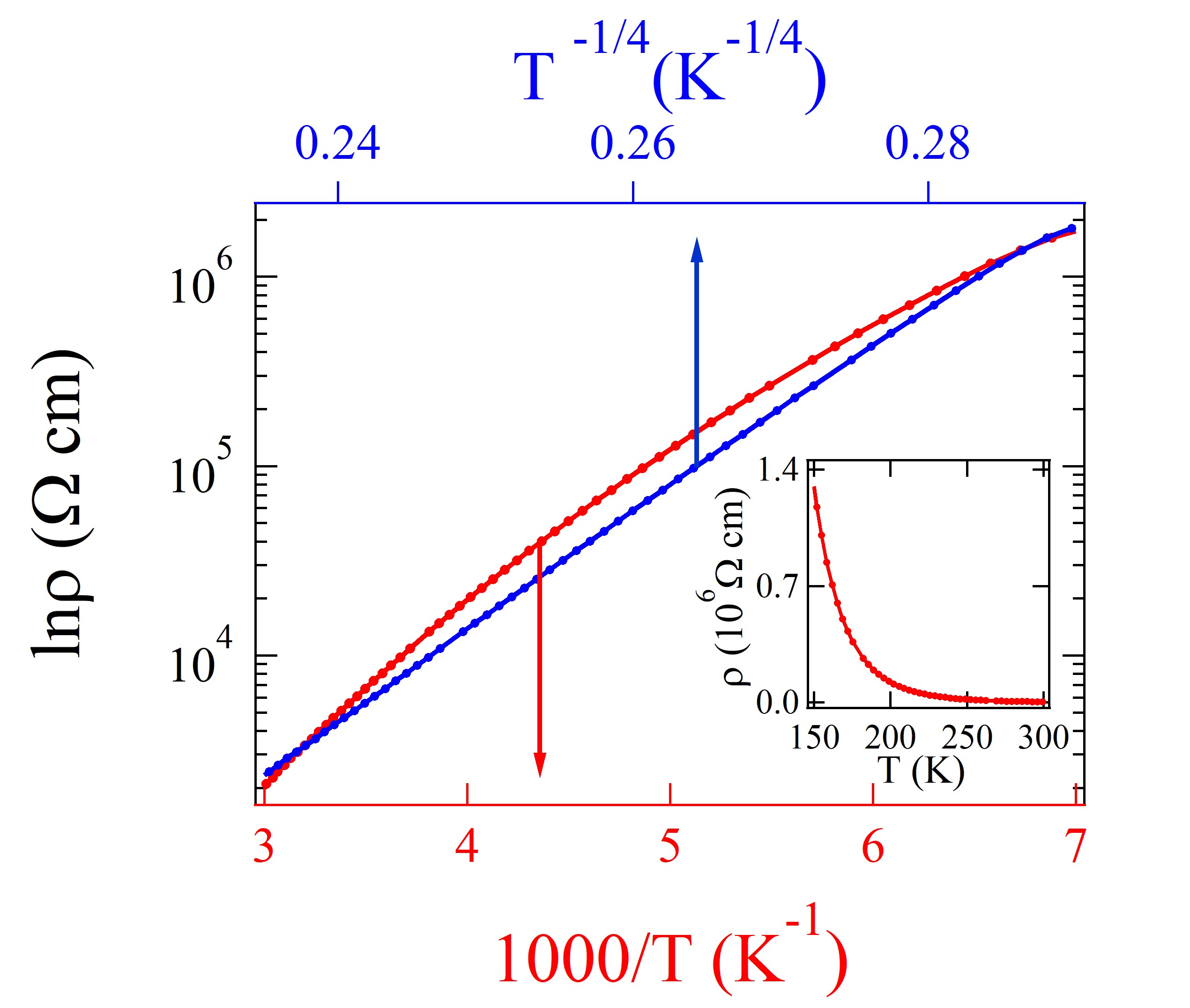}
\caption{\label{Fig2} Electrical resistivity (ln $\rho$) as a function of T$^{-1}$ and T$^{-1/4}$ (150 to 300 K) displayed on the bottom and top axes, respectively. The corresponding plot of resistivity vs temperature ($\rho$ vs T) is shown in the inset.} 
\end{center}
\end{figure}

To know about the dc-electrical resistivity and its change as a function of temperature, we have recorded the dc-resistivity data in the absence of an externally applied magnetic field using PPMS (Quantum Design). Figure \ref{Fig2} shows the plots of the logarithmic variation of $\rho$ (electrical resistivity) as a function T$^{-1}$ and T$^{-1/4}$ for Ba$_{2}$ScRuO$_{6}$ (in the bottom and top panel respectively). The material shows an activated conductivity within the studied temperature zone, indicating a semiconducting-type behavior (inset in Fig. \ref{Fig2}). However, as we decrease, the temperature resistivity starts increasing. Below 150 K, it exceeds the measurement limit of the PPMS (Quantum Design). In the measurement limit (150 K - 300 K), the slope of the $\rho$(T) plot is not constant but varies with temperature, suggesting the Mott variable range hopping (VRH)-type conductivity. This is indicated by a roughly linear response of ln $\rho$ (T) plotted on the T$^{-1/4}$ scale. However, the ln $\rho$(T) vs T$^{-1}$ curve does not show a linear behavior. The linear response of ln $\rho$(T) plotted in T$^{-1/4}$ scale is analogous to the 3D-VRH transport model and is also realized for other disordered double perovskites, for instance, in Ca$_{2}$CoOsO$_{6}$ \cite{43a}. Therefore, the electron conduction corresponds to the Mott 3D-VRH model. The carriers get localized due to the disorder. 

 \begin{figure}[h!]
\begin{center}
\includegraphics[width=1.0\columnwidth]{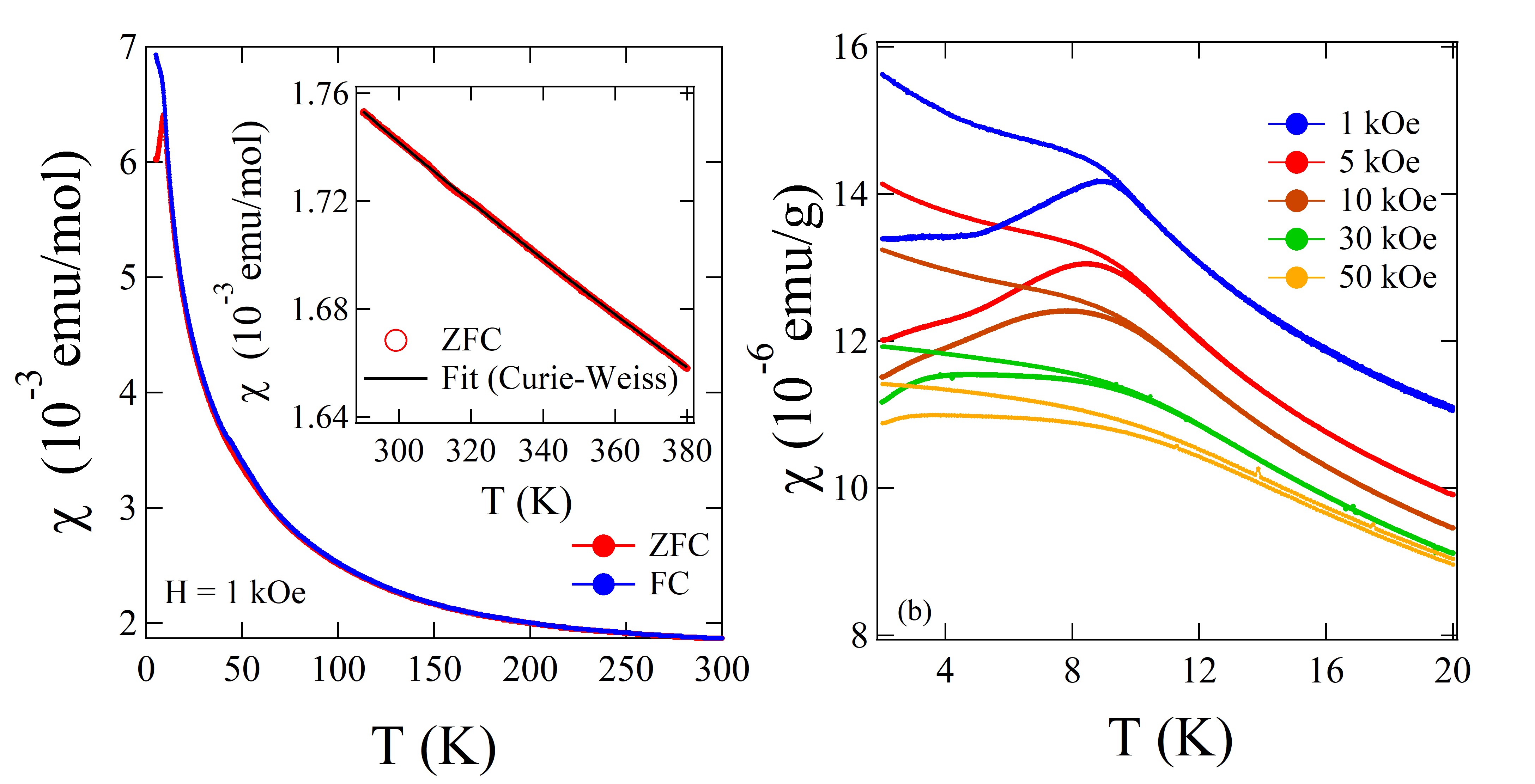}
\caption{\label{Fig3} (a) Temperature-variation of magnetic susceptibility (ZFC and FC, $H$ = 1 kOe) for Ba$_2$ScRuO$_6$. (b) Temperature-dependence of magnetic susceptibility curves (FC and ZFC) in different applied fields.}
\end{center}
\end{figure}

 To understand the magnetic properties of this double perovskite compound, detailed magnetic measurements (magnetisation as a function of temperature and magnetic field) were performed using an MPMS-3 magnetometer. The magnetic susceptibility measurements were done after cooling the sample from RT to 2 K in a zero field cooled (ZFC) and field cooled (FC) mode and under several magnetic fields. The temperature dependence of ZFC and FC susceptibility ($\chi$- T) measured at an applied magnetic field H = 1 kOe for Ba$_{2}$ScRuO$_{6}$ is displayed in figure \ref{Fig3}(a). The ZFC and FC curves are essentially indistinguishable up to $\sim$ 9K. A slight divergence appears below 9 K, showing a magnetic transition. In the ZFC mode, the magnetization shows a pronounced maximum at $\sim$ 9K, signalling a low temperature antiferromagnetic phase due to the magnetic ordering of the Ru$^{5+}$ moments as Sc$^{3+}$ cations are nonmagnetic. The irreversibility between the ZFC and FC curves may arise due to weak ferromagnetic components, possibly attributed to the frustration between the ordered moments.

 \begin{figure}[h]
\begin{center}
\includegraphics[width=1\columnwidth]{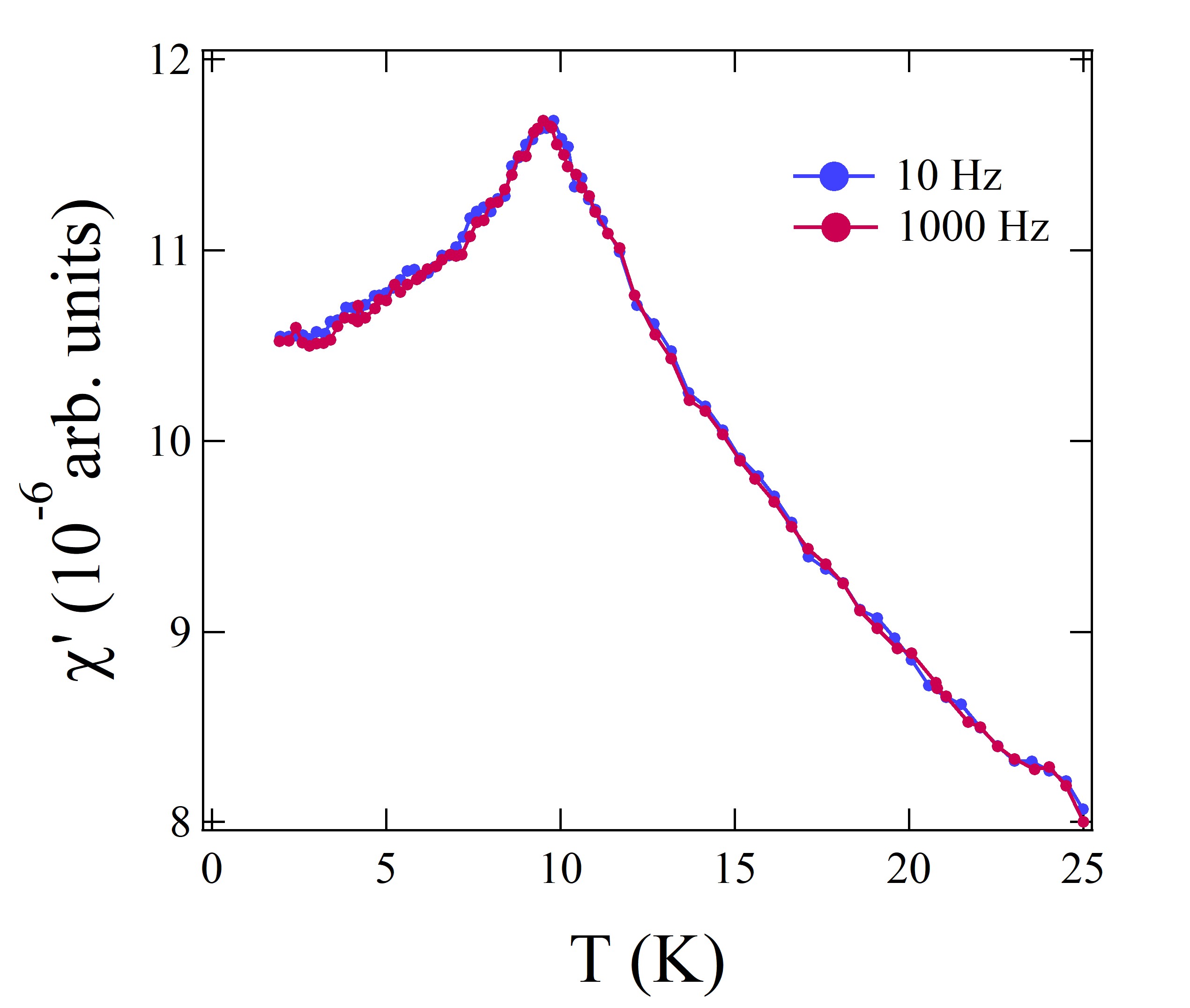}
\caption{\label{Fig4} The temperature dependent real part ($\chi^{\prime}$) of the ac-susceptibility data for Ba$_2$ScRuO$_6$.}
\end{center}
\end{figure}

After normalizing to the measuring field under several applied fields, the ZFC and FC susceptibility plots are illustrated in Fig. \ref{Fig3}(b). The ordering temperature is shifted to lower temperatures for higher applied fields ($>$ 1 kOe). The modified Curie - Weiss fitting  

$\chi = \chi_{0} + \frac{N\mu_{B}^{2}p_{\mathrm{eff}}^{2}}{3k_{B}(T-\theta_{p})}$
 
 [where p$_{eff}$ is the effective paramagnetic moment, $\theta_{p}$ is the paramagnetic Curie temperature, and $\chi_{0}$ is the temperature independent term] in the paramagnetic region above 280 K is shown in the inset of Fig. \ref{Fig3}(a). The black line represents the best fit (inset in Fig. \ref{Fig3}(a)) to the modified Curie-Weiss equation, which agrees with the experimental data in the high-temperature interval. From the fitting between T = 290 K to 380 K, we obtained Curie constant C = 1.12 emu mol$^{-1}$ K and the Weiss temperature $\theta$ = -694 K. The negative value indicates the strong antiferromagnetic correlations between Ru$^{5+}$ ions. Further, the obtained Weiss temperature is very high and indicates the frustration of the magnetic moments at high temperatures. The effective magnetic moment (p$_{eff}$), obtained using C is 2.98 $\mu_{B}$, which slightly differs from the theoretical value for the Ru$^{5+}$ ion (spin-only) with S = 3/2 $\mu_{B}$ (3.87 $\mu_{B}$). However, the calculated value of p$_{eff}$ is similar to those observed for other hexagonal perovskites containing Ru$_{2}$O$_{9}$ dimers, such as Ba$_{3}$YRu$_{2}$O$_{9}$ and Ba$_{3}$LaRu$_{2}$O$_{9}$ compounds \cite{38c}. This can be ascribed to high spin frustration ($f$ = 77).

\begin{figure}[h!]
\begin{center}
\includegraphics[width=1\columnwidth]{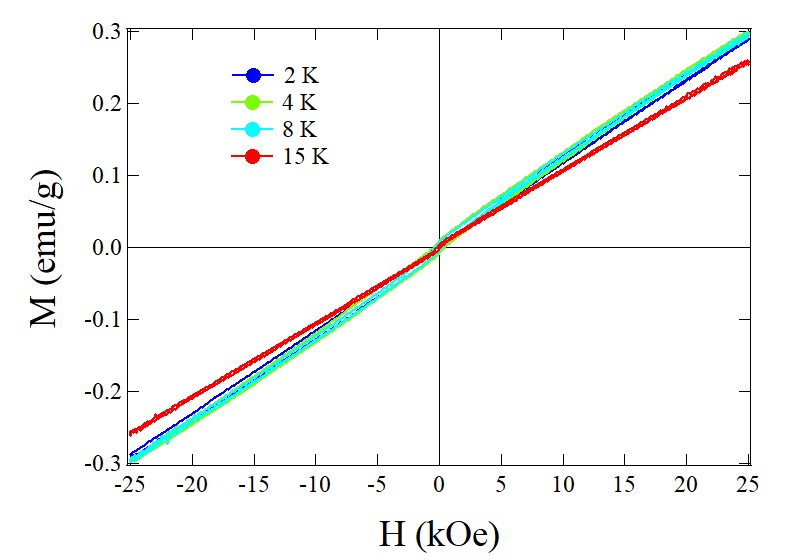}
\caption{\label{Fig5} Zero field cooled magnetization  curves as a function of magnetic field (M - H) for Ba$_{2}$ScRuO$_{6}$ measured at several temperatures.}
\end{center}
\end{figure}

 The $\chi$ - T plots in the presence of several applied magnetic field is shown in \ref{Fig3}(b). The susceptibility values show a decreasing trend with an increasing magnetic field near the magnetic ordering temperature. The susceptibility curves measured in low magnetic fields below T$_N$ = 9 K show high irreversibility between the ZFC and FC modes. However, increasing the applied magnetic field is found to suppress the irreversibility. The high irreversibility (ZFC and FC susceptibility) can also indicate a spin-glass-type transition. However, to get deeper insights into the magnetic state of Ba$_2$ScRuO$_6$, we have performed ac susceptibility measurements at different frequencies (10 Hz and 1 kHz). Fig. \ref{Fig4} shows the real part ($\chi^{\prime}_{ac}$) of our ac susceptibility measurements. The peak position in the ac susceptibility curves does not show a frequency-dependent shift, confirming the absence of a spin-glass-type behavior in Ba$_2$ScRuO$_6$. To further understand the magnetic behavior in detail, magnetic isotherms ($M$-$H$) were obtained between $\pm$25 kOe at different temperatures (between 2 K - 15 K) in zero field cooling mode (figure \ref{Fig5}). A linear change in magnetization (AFM type) with very small hysteresis behavior is observed. It could be due to the existence of frustration in the structure.

\begin{figure}[h!]
\begin{center}
\includegraphics[width=1\columnwidth]{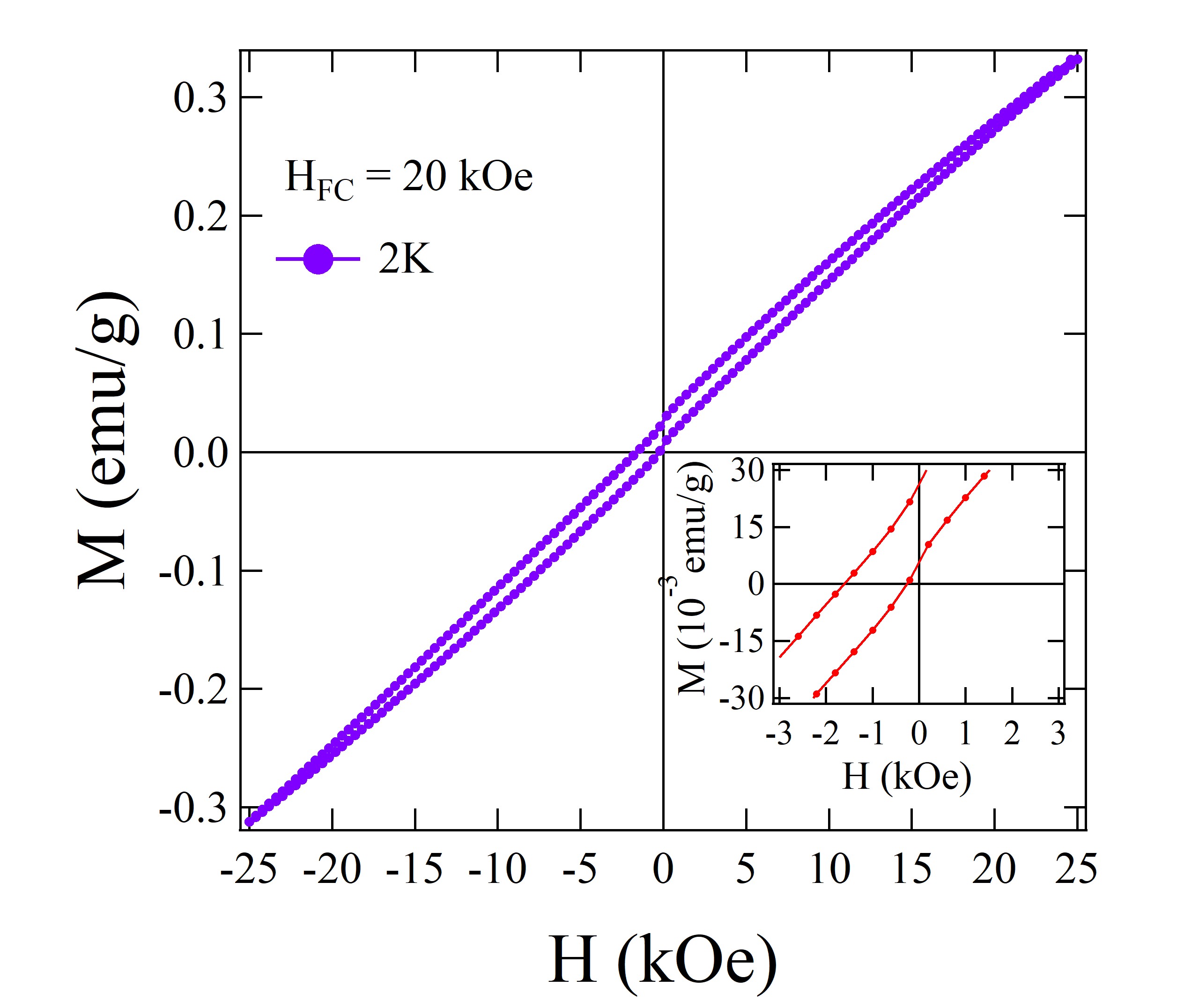}
\caption{\label{Fig6} Field cooled ($H_{FC}$ = 20 kOe)) $M$-$H$ loop for Ba$_2$ScRuO$_6$ measured at 2 K. The inset illustrates the enlarged central part of the FC $M$-$H$ loop.}
\end{center}
\end{figure}

Recently, there has been a significant interest in exploring and understanding exchange bias (EB) effects in structurally inhomogeneous systems with various combinations of magnetic cations \cite{18l, 182, 183}. In exchange bias effect, the field cooled (FC) M - H loops show a shift along the magnetic field axis, and this is used in several technological applications such as magnetic recording read heads \cite{30l}, random
access memories \cite{31l}, and other spintronics devices \cite{32l,33l}. To explore the behavior of EB in the present sample, we have measured a $M$-$H$ loops at selected temperatures in the field cooling condition ($H_{FC}$ = 20 kOe). The $M$-$H$ loops were measured in the FC mode between $\pm$25 kOe. For all the M - H measurements (ZFC and FC), the material was cooled from T $>$ T$_{N}$ to measuring temperature (T$_{meas}$) and the $M$-$H$ data are obtained in the field range $\pm$ 25 kOe when reaching T$_{meas}$. The FC M-H curves measured at 2 K is shown in Fig. \ref{Fig6}. Interestingly, the field-cooled M - H loops show a shift along the negative field axis. A representative result, as illustrated in Fig. \ref{Fig6}, highlights a shift of the center of the M - H curve along the negative magnetic field axis of around 1.24 kOe. Being a classic signature of unidirectional anisotropy, the shift of the magnetization loop ($M$-$H$) reveals the surprising existence of the EB effect in Ba$_{2}$ScRuO$_{6}$. The offset of the M - H curve along the field axis can be denoted as exchange bias field $H_{EB}$ and the coercive field $H_{C}$, and can be extracted from the M - H loops [$H_{C}$ = $(H_{+} - H_{-})$/2 and $H_{EB}$ = $(H_{+} + H_{-})$/2, respectively. Here, $H_{+}$ and $H_{-}$ are the coercivities in the ascending and descending branch of the M - H curve with field axis]. 

\begin{figure}[h!]
\begin{center}
\includegraphics[width=1.0\columnwidth]{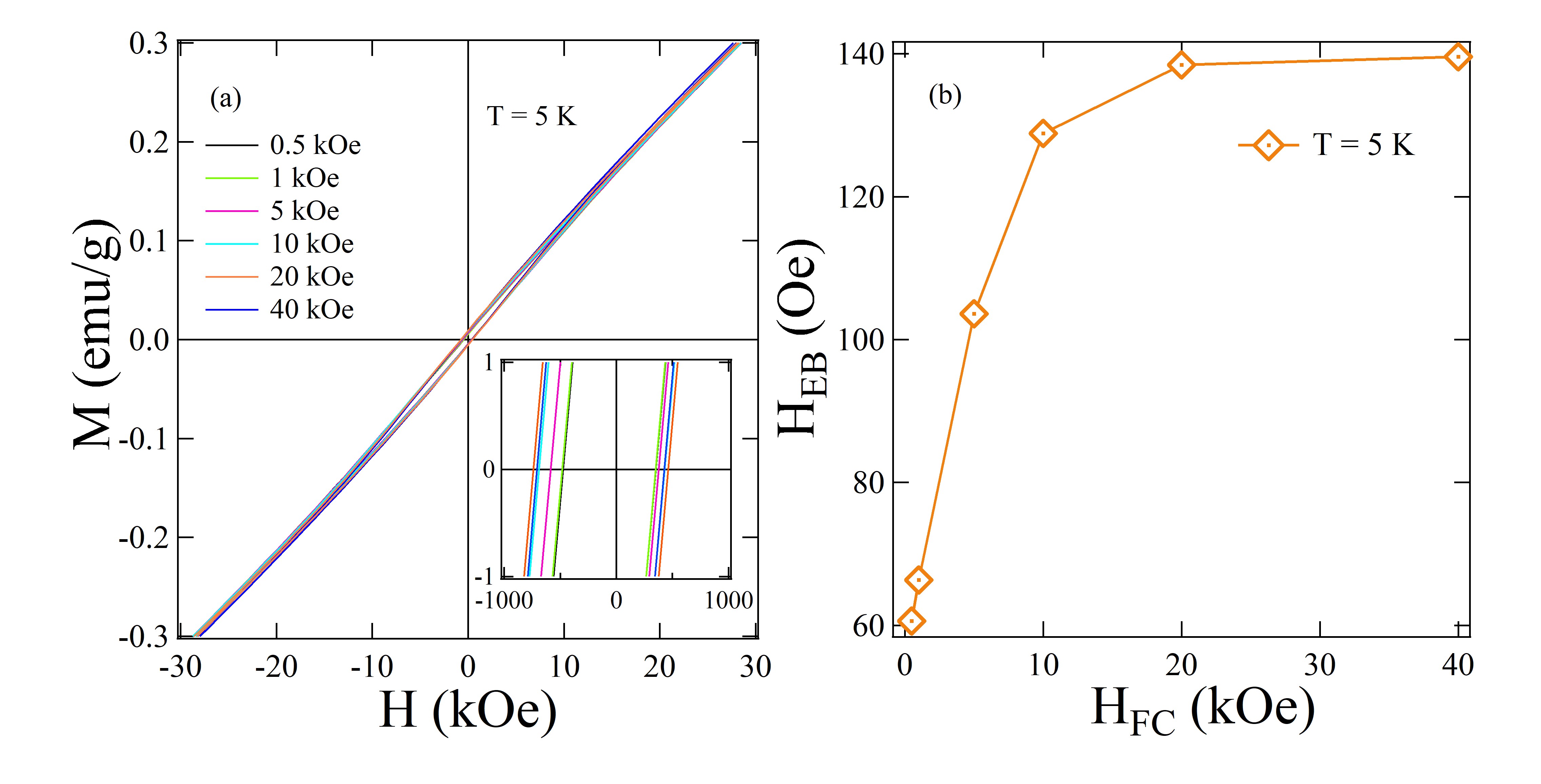}
\caption{\label{Fig7}(a) $M$-$H$ loops measured at T = 5 K with different values of $H_{FC}$ (0.5 kOe to 40 kOe). The inset highlights the central and enlarged part of the $M$-$H$ loops. (b) Variation of $H_{EB}$ with different $H_{FC}$ values (0.5 kOe to 40 kOe).}
\end{center}
\end{figure}

The choice of cooling field is extremely important in exchange bias measurement. To choose a suitable $H_{FC}$ for Ba$_{2}$ScRuO$_{6}$, the $M$-$H$ loops are obtained at 5 K while cooling the sample in a different magnetic field ranging from 0.5 to 40 kOe and are depicted in Fig. \ref{Fig7} (a). $H_{EB}$ increases with the cooling field up to 10 kOe, and then it exhibits a saturation tendency for $H_{FC}$ above 10 kOe [see Fig. \ref{Fig7} (b)]. Therefore, we have considered a cooling field 20 kOe (more than 10 kOe) to study the EB effect in Ba$_{2}$ScRuO$_{6}$. 

\begin{figure}[h]
\begin{center}
\includegraphics[width=1\columnwidth]{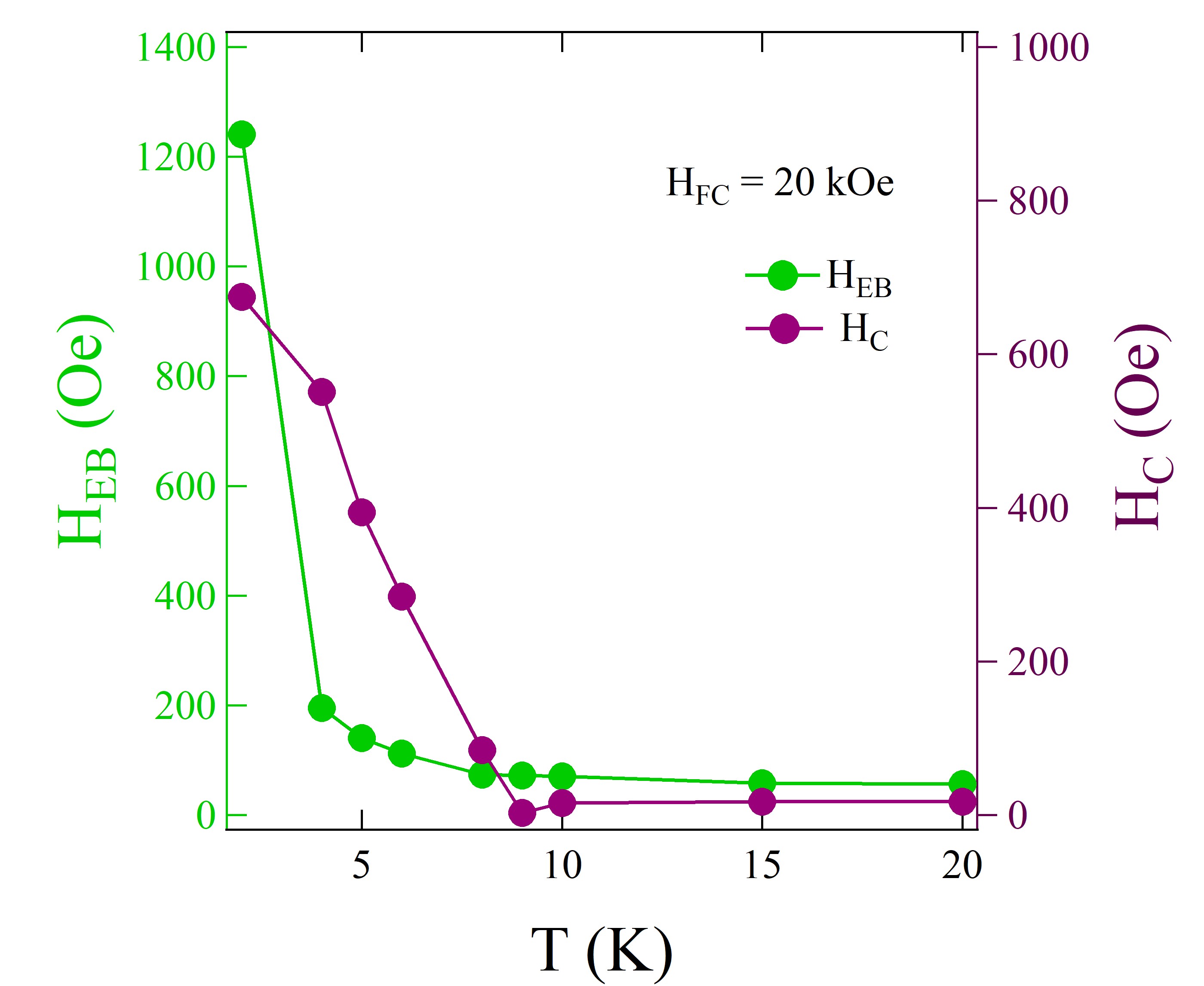}
\caption{\label{Fig8} Left axis: Temperature-dependent $H_{EB}$ for Ba$_2$ScRuO$_6$. The $H_{EB}$ values are obtained from the FC M - H curves measured at several temperatures. Right axis: Temperature-dependent $H_{C}$ obtained from the same M - H loops.}
\end{center}
\end{figure} 

It is important to know whether the exchange bias properties are associated with long-range magnetic ordering in Ba$_{2}$ScRuO$_{6}$. Therefore, we have measured several $M$-$H$ curves at various fixed temperatures. The characteristic temperature-dependent trend of the exchange bias field and coercivity with the measuring temperature for Ba$_{2}$ScRuO$_{6}$ is summarized in Fig. \ref{Fig8}. A smooth decrease in the exchange bias field ($H_{EB}$) and the coercivity ($H_{c}$) is observed with increasing temperature, eventually disappearing above $T_{N}$ = 9 K. Therefore, the appearance of an exchange bias field and coercivity below the magnetic ordering temperature indicate its relation with the magnetic correlations in the structure. Besides the long-range AFM ordering ($T_{N}$ = 9 K), the disorder in the Ru - Sc layers could originate from magnetic frustration (Weiss temperature = -694 K, f = 77). The high magnetic frustration in the structure, therefore, can cause an exchange bias effect below $T_{N}$ = 9 K.\\

\begin{figure}[h]
\begin{center}
\includegraphics[width=1.0\columnwidth]{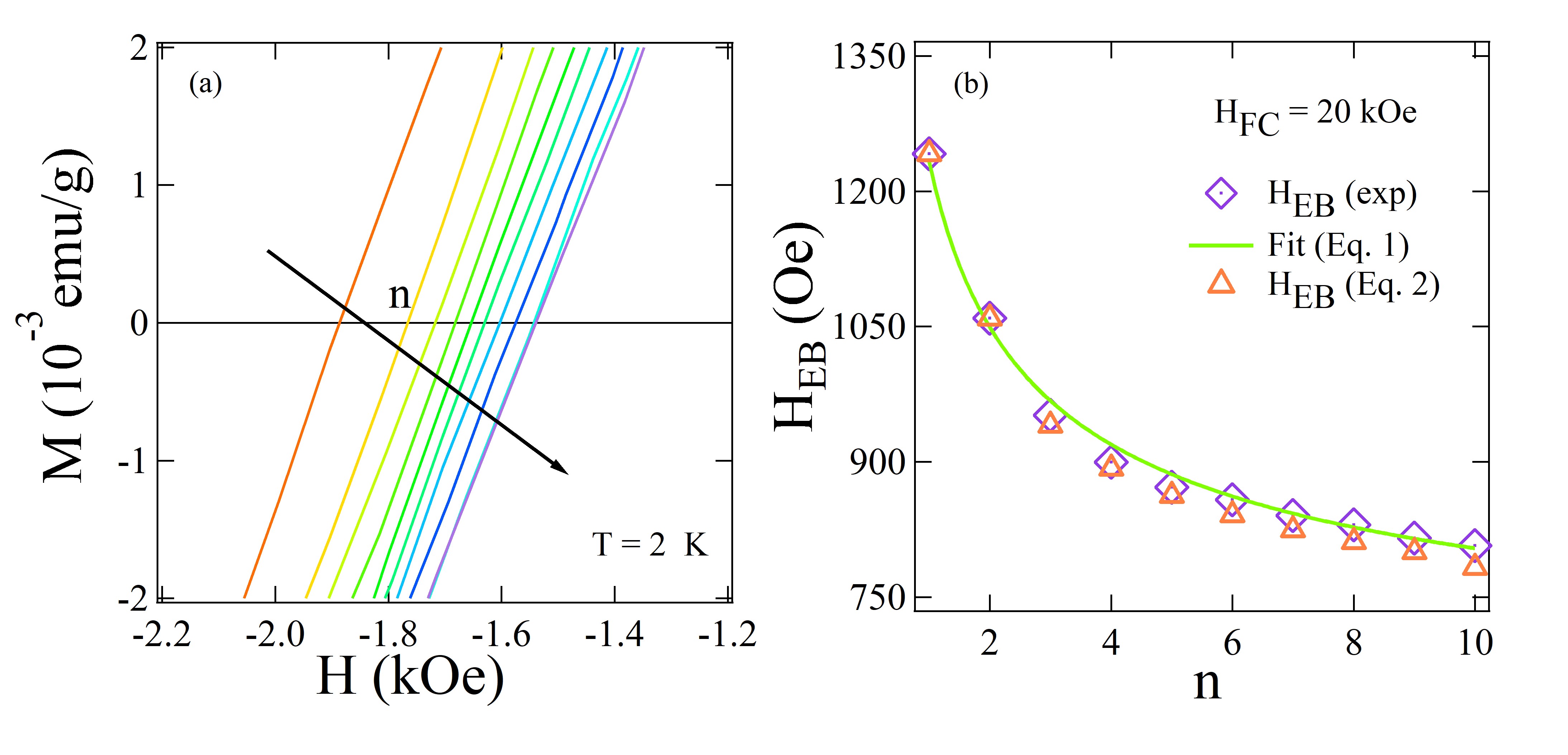}
\caption{\label{Fig9}(a) The expanded view of the central part of the consecutive $M$-$H$ loops. (b) The dependence of $H_{EB}$ on the number of field cycles ($n$) (diamond markers) obtained from the consecutive $M$-$H$ loops (training effect). The green line is the best fit to Eqn. \ref{eqn1:h} for $n >$ 1. Open triangles (orange) data points are generated using Eqn. \ref{eqn2:hm}.}
\end{center}
\end{figure}

	The uncompensated spin configuration of the present system can relax from its equilibrium configuration in the presence of suitably applied magnetic fields. Therefore, the system is trained by increasing the number of successive hysteresis loops ($M$-$H$). In these connections, the training effect is an important property, generally highlighted in EB materials. This reduces the exchange bias field H$_{EB}$ as the M -H curve shrinks when the material is cycled in several consecutive $M$-$H$ loops. Figure \ref{Fig9} (a) shows the enlarged version of the central portion of the representative magnetic loop recorded at 2 K in the negative field quadrant after the field cooled to 20 kOe. A prominent training effect is observed here.  H$_{EB}$ as a function of the loop cycles (n) is shown in \ref{Fig9} (b). The training effect is generally considered a direct macroscopic fingerprint of the rearrangements of the spins toward equilibrium [see Fig. \ref{Fig9}] \cite{a94, CHB}. The training effect gets prominent after the first magnetic field cycle due to the reconfiguration of the frozen spins. The reconfiguration in the spin structure results in a sharp decrease in H$_{EB}$ after the first magnetic field cycling. Consequently, the change in $H_{EB}$ after the second field cycle is moderate.
The usual relationship between H$_{EB}$ and $n$ (for $n >$ 1) that describes the relaxation of the EB effect is known to lead to a simple power-law relationship \cite{BattleP};
\begin{equation}
H_{EB}(n)-H_{E\infty} = k/\sqrt{n}  ,
\label{eqn1:h}
\end{equation}
Here H$_{EB}$ (n) = exchange bias field at the nth cycle, H$_{E\infty}$ = exchange field in the limit $n \rightarrow \infty$ and $K$ = system dependent constant. The best-fitted results (dotted line) are shown in Fig. \ref{Fig9} (b). The value of H$_{E\infty}$ obtained from the fitting is 606.24 Oe, which is the remnant H$_{E\infty}$ in the material. 

Furthermore, an additional approach is required to explain the training effect since the power law described in Eq. \ref{eqn1:h} breaks down for $n$ = 1. It was proposed in the framework of non-equilibrium thermodynamics, and it causes a relaxation toward equilibrium in the spin configuration of the interface magnetization (anti-ferromagnetic). To better describe the exchange bias effect and its dependence on the number of field cycles ($n$), a recursive formula is considered \cite{BC2}; 
\begin{equation}
H_{EB}(n+1)-H_{EB} (n) = \gamma (H_{EB}(n)-H_{E\infty})^{3}  ,
\label{eqn2:hm}
\end{equation}                   																							
where, $\gamma$ = sample dependent constant (= 1/2$k^{2}$) and $H_{E\infty}$ = exchange bias field in $n$ $\rightarrow$ $\infty$. Using $\gamma$ = 1.28$\times$10$^{-6}$ Oe$^{-2}$  and $H_{E\infty}$ = 606.24 Oe (generated from the power law fitting, Eq. \ref{eqn1:h}) as an additional input in Eq. \ref{eqn2:hm}, the whole set of results can be generated for $H_{EB}$. We have displayed the obtained results by open (orange) triangles in Fig. \ref{Fig9} (b).

Giant EB effects have recently been reported in a similar double perovskite material with composition Ba$_{2}$Fe$_{1.12}$Os$_{0.88}$O$_{6}$ (having 6L hexagonal structure) \cite{40a}. The presence of antisite disorder originates from competing magnetic interactions and frustration in the system and is attributed to the huge EB effects in these systems. Material systems with different crystallographic sites for magnetic cations are prone to antisite disorder. In the present material, a disorder between Ru - Sc is observed at both the Sc2 (M3, 1a, corner-sharing B$_{2}$O$_6$ sites) and Ru2 (M4, 2d, face-sharing B$_{2}$O$_{9}$ dimer) sites. However, the degree of disorder is less than the Ba$_{2}$Fe$_{1.12}$Os$_{0.88}$O$_{6}$ material. Therefore, similar to the Ba$_{2}$Fe$_{1.12}$Os$_{0.88}$O$_{6}$ material, the disorder in Sc - Ru, which is originating magnetic frustration is responsible for the EB effect in Ba$_{2}$ScRuO$_{6}$ below the AFM ordering temperature.

\section{Conclusions}
In conclusion, the ambient pressure synthesized Ba$_{2}$ScRuO$_{6}$ is found to crystallize in a 6L hexagonal structure (space group P$\overline{3}$m1) at room temperature. Electron conduction corresponds to the Mott 3D-VRH model, and the disorder initiates carrier localization. This antiferromagnetic double perovskite compound Ba$_{2}$ScRuO$_{6}$ shows exchange bias below the long-range antiferromagnetic ordering ($T_{N}$ = 9 K). The exchange bias effect is seen below the AFM ordering temperature when the sample is cooled with a magnetic field ($H_{FC}$ = 20 kOe) and reaches a maximum value of 1.24 kOe at 2 K. In addition, studies of training effects studies support exchange bias. Inhomogeneous magnetic correlations and magnetic frustration formed due to the disorder of magnetic (Ru) and nonmagnetic (Sc) cations at both the Sc2 (1a, corner-sharing B$_{2}$O$_6$ sites) and Ru2 (2d, face-sharing B$_{2}$O$_{9}$ dimer) sites. The magnetic frustrations in the structure below the long-range AFM ordering originate from the EB effect in this sample.

\section*{Acknowledgements}
R. P. S. acknowledges the Science and Engineering Research Board (SERB), Government of India, for the CRG/2019/001028 Core Research Grant. S. M acknowledges the Science and Engineering Research Board, Government of India, for the SRG/2021/001993 Start-up Research Grant.

\end{document}